\documentstyle[11pt,epsf]{article}

\begin{document}
\def\be{\begin{equation}}
\def\ee{\end{equation}}
\def\bc{\begin{center}}
\def\ec{\end{center}}
\def\bea{\begin{eqnarray}}
\def\eea{\end{eqnarray}}

\begin{center} 
\large{ \bf
 Non perturbative renormalization group approach to surface growth
}
\end{center}

\begin{center}
{\noindent  M. A.  Mu{\~{n}}oz$^{(1)}$, G. Bianconi$^{(1)}$,
 C. Castellano$^{(2)}$
  A. Gabrielli$^{(1,3)}$,  \\ M. Marsili$^{(4)}$,
and L. Pietronero$^{(1,2)}$}
 \end{center}
\begin{center} 
{\noindent $^{(1)}$
 Dipartimento di Fisica and INFM Unit,
University of Rome ``La Sapienza'', I-00185 Roma\\
$^{(2)}$
 The ``Abdus Salam'' I.C.T.P.
Strada Costiera 11, I-34100 Trieste\\
$^{(3)}$ University of Rome, ``Tor Vergata'', I-00133 Roma\\
$^{(4)}$
International School for Advanced Studies
(SISSA), via Beirut 2-4, Trieste I-34014}
 \end{center}

\begin{abstract}
We present a recently introduced real
 space renormalization group (RG) approach
to the study of surface growth.
 The method permits us to obtain 
the properties of the KPZ strong coupling fixed point,
 which is not accessible
to standard perturbative field theory approaches.
Using this method, and with the aid of small Monte Carlo
 calculations for systems 
 of linear size $2$ and $4$, we calculate the roughness exponent 
in dimensions
up to d=8. The results agree with the known numerical values
 with good accuracy.
Furthermore, 
 the method permits us to predict
the absence of an upper critical dimension
 for KPZ contrarily to recent claims.
The RG scheme is applied to other
 growth models in different universality classes and
 reproduces
very well all the observed phenomenology and numerical results.
Intended as a sort of finite size scaling method, the new scheme
may 
simplify in some cases from a computational point of view 
the calculation of scaling exponents of growth processes.
\end{abstract}


 \vspace{2mm}     
  The study of the non equilibrium dynamics of rough surfaces and
interfaces is a topic of growing interest \cite{HZ,Krug,Laszlo}.
  Many efforts have been
devoted to single out the behaviors shared by apparently
different growing phenomena. In this context
the search of universality 
classes, permitting to categorize different models and systems of
rough surfaces sharing the same scaling properties, is a central task.
  The main milestone in this direction is the Kardar-Parisi-Zhang
equation (KPZ), which
 is the minimal Langevin equation
capturing the     
essential physics of many different
 growing surfaces beyond the Gaussian linear (Edward-Wilkinson)
theory \cite{HZ,Krug,Laszlo}.
It reads  \cite{kpz}
\be
{\partial h(x,t) \over \partial t} = \nu \nabla^2 h + {\lambda \over 2}
(\nabla h)^2 + \eta(x,t).
\ee
where $h(x,t)$ is a  height variable at time t and position $x$ in a
d-dimensional 
substrate of linear size $L$. $\nu$ and $\lambda$ are constants
 and $\eta$ is a Gaussian white noise.
The KPZ equation has become a paradigm of surface growth processes and an 
overwhelming amount of both numerical and analytical studies 
have been performed since it was first proposed. It is also related 
to other physical systems as:
directed polymers in random media \cite{Krug}, systems with 
multiplicative noise \cite{MN},
and the Burgers 
equation \cite{burger}.

A central quantity of interest in the study of surfaces 
 is the roughness $W(L,t)$,     
defined as the fluctuation of the height variable 
 around its mean value,
$ \bar{h} (t)$;
this is:
\be 
 W^2(L,t)=\frac{1}{L^d}\sum_{x} [h(x,t) -\bar{h} (t)]^2
\ee 
   In many seemingly unrelated problems the large scale behavior
of the roughness is observed
to be scale invariant, in the sense that it grows in time as
 $W(L,t) \sim t^\beta$ until it saturates to a stationary value
 at a characteristic
time, $t_s \sim L^z$, with universal exponents $\beta$ and $z$.
 Once in the saturated regime, the roughness 
obeys 
$W(L) \sim L^\alpha$.
These critical exponents are not independent, in fact, from a trivial
scaling argument $\alpha = \beta z$, and as a consequence
of a tilting invariance 
\cite{burger,HZ,Krug} $\alpha + z=2$. There is therefore only
one independent exponent, say $\alpha$.

Standard field theoretical RG studies of the KPZ equation
 predict the presence of a
{\it roughening transition } above $d=2$ \cite{Terry}; i.e. there are two RG attractive fixed points:
A trivial (Gaussian) one with $\alpha=0$ which describes a flat phase,
 and one describing the
rough phase. This last one is {\it non-perturbative}, therefore 
standard perturbative methods fail to give any prediction 
for the exponents in the rough phase.
One fundamental question, presently under debate,
 is to elucidate if there is a finite upper
critical dimension
\cite{debate}. Summing up, any new  systematic approach to the problem of surface growth
  would be certainly desirable.

In this paper we review a recently introduced real-space, non-perturbative
renormalization group scheme \cite{First,second,Ginestra}
 permiting to determine roughness
exponents in growth processes (even in strong coupling regimes) with 
the aid of relatively small Monte Carlo computations.
This method can be seen as a finite size
scaling method, permitting to determine from the
analysis of very small system sizes (which is very 
convenient from a computational point of view)
 scaling exponents with reasonable
accuracy, and constitutes  a powerful tool in the analysis
of growth processes.

Let us now describe the method 
 as applied to a microscopic model
 in the KPZ universality class \cite{First}:
 The solid-on-solid (SOS) model (see \cite{HZ,Krug,Laszlo} for the definition and details).
  Instead 
 of studying its microscopic
 dynamics defined over cells of unit size $L_0=1$, and height $h_0$
 in a substrate of
total length $L$, we describe the model at a coarse-grained level
 \cite{First}. 
For that purpose we first define blocks or cells, of linear
size  $L_k=2^k$ and height $h_k = h_0 {(2^K)}^{\alpha}$. 
$\alpha$ is for the time being
 an unknown exponent to be determined  latter in a self-consistent
way.
  A system configuration is specified
by giving the heights $h(i)$ at scale $k$,
 at every substrate site, $i$ ($i=1,..., b$,
where $b= L/2^k$). Two configurations are considered
equivalent if they differ by a constant height at every site.
 Having defined the blocks, 
now  we introduce an 
effective coarse-grained dynamics, or growth rules at a generic scale. 
This is done by giving 
the transition rates from a generic surface configuration
to a new one with one additional grown block.
The transition 
rates should contain the 
relevant information of the microscopic dynamics.
The growing rate at site $i$ is defined as
\be
P[h(i)\to h(i)\!+\! h_k] \equiv L_k\!+\!\lambda_k \!
\sum_{j nn i}\max[0,h(j)-h(i)].
\label{rates}
\ee
The term $L_k$ is the contribution of the vertical growth (i.e. 
random deposition) and the
sum over neighbor block sites $j$ accounts for the lateral
growth.  We hypothesize that these are the two only possible types of growth
in the 
dynamics of the SOS
coarse-grained dynamics. 
The ratio  $x_k=\lambda_k/L_k$ represents the relative importance 
of the lateral growth from an elementary lateral step in comparison with the 
growth through random deposition at scale $k$.
 The parameter $x_k$, necessary to fully
specify the dynamics is
to be determined self-consistently
by imposing scale invariance. 

  Iterating in time the aforementioned growth rules, a statistically 
stationary 
state should be reached, the averaged width of which could be determined 
as a function of $L_k$, $x_k$ and $h_k$. 
  At this point we can pose ourselves the following question:
given a description of the system at scale $k$, defined by
parameters ($L_k$, $h_k$, $x_k$),
what will be the values of $h_{k+1}$ and $x_{k+1}$
at the scale $L_{k+1}$?
To answer this question we consider a given system of size
 $L=L_{k+2}$ and describe it at
two different levels of coarse graining: using $b$ blocks of size
$L_{k+1} = L_{k+2}/b$ and then dividing each component block of size $L_{k+1}$
again in $b$ blocks of size $L_k$, or alternatively considering directly $b^2$ blocks of
size $L_k = L_{k+1}/b = L_{k+2}/b^2$. 
The condition that the 
total roughness
takes the same value when calculated using blocks of two different scales 
gives a renormalization equation
 in a way to be explained in what follows. In particular,
we use the following a property of 
  the roughness, $W^2$:
 Partitioning the system in cells
of size $a$, one obtains
$W^2(L)= W^2(a)+ W^2(L/a)$,
 i.e. the total roughness, $W^2(L)$, is equal to the roughness of the
system measured using cells of size $a$, $W^2(L/a)$, plus the 
internal roughness of one of these cells, $W^2(a)$.
  First we apply the last equation to the case in which $L=L_{k+1}$ and
$a=L_k=L_{k+1}/b$, obtaining:
$W^2(L_{k+1})=W^2(L_{k})+W^2(b;x_k)$
where $W^2(b;x_k)$ is the stationary roughness for $b$ blocks 
for $x_k$, which 
  is proportional to $h^2_k$,
and therefore can be formally written as $W^2(b;x_k) =  R(x_k,b) h^2_k$.
 As shown in \cite{First} $h^2_k=c*W^2(L_k)$ where $c$ is a constant
\cite{scel}; this simply reflects the fact
 that the height $h_k$ at scale $k$ has
to be taken proportional to the roughness at that scale.
Combining the last two results, we can write
\be
W^2(L_{k+1}) = W^2(L_{k}) ( 1 + c R(x_k,b)).
\label{f1}
\ee
Applying twice this relation we can relate scales $k+2$ and $k$,
\be 
W^2(L_{k+2}) = W^2(L_{k}) ( 1 + c R(x_{k+1},b)) (1+ c R(x_k,b)).
\label{f2}
\ee
Another possibility, as said previously, is to relate directly scales $k+2$ and
$k$, applying once the relation Eq.(\ref{f1}) replacing 
$b$ by $b^2$. This leads to:
\be
  W^2(L_{k+2}) = W^2(L_{k}) ( 1 + c R(x_k,b^2)).
\label{f3}
\ee
  Equating the r.h.s. of Eq.(\ref{f2}) and Eq.(\ref{f3})
we get the renormalization equation:
\be 
( 1 + c R(x_k,b^2)) = ( 1 + c R(x_{k+1},b)) (1+ c R(x_k,b))
\label{f4}
\ee
which gives, once the function $R$ is known, 
 the flow of the renormalized parameter $x_k$ 
under changes of scale, 
and
therefore permits one to evaluate the 
fixed point value $x^*$ that characterizes the scale invariant 
dynamics. 
 Knowing $x^*$, it is a straightforward task to determine 
the roughness exponent $\alpha$:
\be
\alpha = \lim_{k \rightarrow \infty}
 \frac{1}{2} \log_2  \left(
\frac{W^2(L_{k+1},x^*)}{W^2(L_{k},x^*)}  \right) =  \frac{1}{2}\log_2 
( 1 + c R(x^*,b)).
\label{f5}
\ee
 
 The previously exposed method  was applied in \cite{First}
 to study KPZ growth in
finite dimensions (from $d=1$ to $d=8$)  by analyzing the case $b=2$.
The saturated width (i.e. the function $R$)
 was evaluated through Monte Carlo simulations
of the effective dynamics
 in these small cells (namely $2^d$ and $4^d$)
for different values of $x$.
In all dimensions a stable fixed point is found with the associated  values of
$\alpha$
 reported in table (1).

\begin{table}
\begin{center}
\begin{tabular}{c|cccccccc}
$d$                & 1    & 2    & 3    & 4    & 5    & 6    & 7     & 8    \\
\hline
$\alpha_{\rm RG}$  & 0.502& 0.360& 0.284& 0.238& 0.205& 0.182& 0.162 & 0.150 \\
$\alpha_{\rm num}$ & 0.5  & 0.387& 0.305& 0.261& 0.193& 0.18 & 0.15  &  -   \\
\hline
\end{tabular}
\caption{Results of the renormalization group calculation 
 (first row) compared with numerical results (second row).}
\end{center}
\label{tabella}
\end{table}

Observe the good agreement between numerical results obtained 
after extensive simulations  \cite{Ala}, and our estimations using 
the new RG scheme with small cells \cite{First}.
 These results support both the validity of the numerical 
simulations \cite{Ala} and of our RG method.
 In particular, for dimensions larger
than $d=4$ exponents differing from the mean field prediction $\alpha=0$
are observed, providing evidence that $d=4$ is not
the upper critical dimension of KPZ. 

 The method can be applied to study analytically the infinite 
dimension limit \cite{second}. For that, one just has to evaluate 
the width by using some theoretical argument. Assuming that 
the lateral growth rate is very large for large dimensions  
(assumption that can be verified a posteriori),
it is a good approximation
(as discussed in \cite{second})
to consider only surfaces with two layers: high and low respectively. 
A state of the system can be characterized by the number of sites 
in high (low) position, and a master equation giving the flow
between different system states can be written down. Calculating
analytically 
the width of its associated stationary state with the previous
hypothesis for two different 
cell sizes, it is a matter of simple algebra to obtain a lower 
bound for $\alpha$ for large $d$ \cite{second}
\be
\alpha \ge   \frac{1}{3(\ln 2)^2}\frac{1}{d}.
\ee
and the associated fixed point is stable for all $d$. This result
excludes the possibility of having a finite upper critical dimension 
for KPZ-like processes.

   Even though we have presented the method as a scheme to analyze KPZ-like
processes the underlying ideas are quite general and can in principle
 be applied
to other growth processes. In particular we have studied linear 
growth models (i.e. KPZ with $\lambda=0$), that should belong in the 
free (Gaussian) universality class, and verified that, as expected 
\cite{Laszlo},
the non-trivial fixed point seems to disappear for $d \ge 2$. We have also
applied the method to other highly non-trivial growth models that exhibit
roughening transitions even in one dimension \cite{Evans}.
 In particular we have constructed a two-parameter representation 
of the model presented in \cite{Evans}: with probability $p$
 a particle is added at a randomly chosen site,
while with complementary probability $1-p$ the value of the height 
at that site is changed by averaging it with its nearest neighbors
weighted with a parameter $a$ \cite{Ginestra}.  This model exhibits a rich
phenomenology as the parameters, $a$ and $p$  are varied \cite{Ginestra}.
There are three different scaling regimes:
(i) $\alpha=1$ for $a=1$, (ii) for $a < 1$ and
small values of $p$ there is a  smooth phase with $\alpha=0$, and
(iii) for  $a < 1$ and larger values of $p$, $\alpha=0.5$.
This complex behavior is perfectly reproduced when the model 
is analyzed the RG scheme. In particular in figure 1, we show the 
RG flow lines in the two-dimensional parameter space showing the
presence of different fixed points with the expected values
of $\alpha$, and a separatrix signaling the roughening transition.        

   All the previous results support the validity of the
 new non-perturbative real space RG scheme as an useful tool 
to analyze growing surfaces. The method itself can be considered
as a sort of finite-size scaling procedure, and  provides therefore a 
computational tool to facilitate the computation of scaling
exponents.

\begin{figure}
\centerline{
\epsfxsize=2.5in
\epsfbox{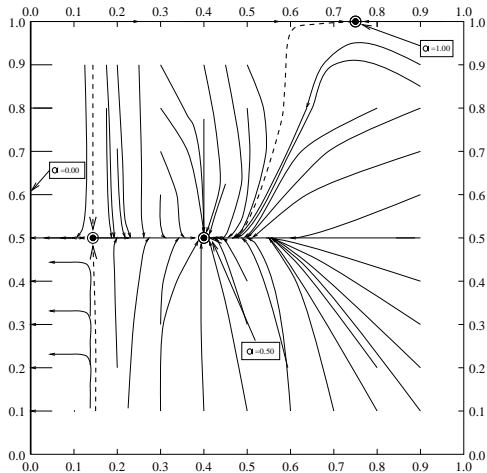}}
\caption{ RG flow for the two parametric model described in the text. 
}
\label{fig1}
\end{figure}


We acknowledge interesting discussions with A. Maritan, G. Parisi,
A. Stella, C. Tebaldi and A. Vespignani. This work has been partially 
supported by the European network contract FMRXCT980183, and by 
a M. Curie fellowship,    
ERBFMBICT960925, to M.A.M.


\begin{thebibliography}{99}
\bibitem{scel} The more natural choice for $c$ is $4$; see \cite{First}.
\end{thebibliography}
\end{document}